\documentclass[namedreferences]{solarphysics}
\usepackage[optionalrh]{spr-sola-addons} 
\usepackage{graphicx}        
\usepackage{color}           
\usepackage{url}             
\usepackage{subfig}

\begin{document}

\begin{article}

\begin{opening}

\title{Uncovering the Birth of a Coronal Mass Ejection from Two-Viewpoint SECCHI Observations}

\author{A.~\surname{Vourlidas}$^{1}$\sep
        P.~\surname{Syntelis}$^{2,4}$\sep
        K.~\surname{Tsinganos}$^{3,4}$
       }
\runningauthor{Vourlidas et al.}
\runningtitle{Uncovering the Birth of a CME}

\institute{
$^{1}$ Space Sciences Division, Naval Research Laboratory, Washington DC, USA \\
$^{2}$ Research Center for Astronomy and Applied Mathematics, Academy
of Athens, Athens, Greece \\
$^{3}$ National Observatory of Athens, Athens, Greece\\
$^{4}$ Section of Astrophysics, Astronomy and Mechanics, Department of
Physics, University of Athens, Athens, Greece\\
              }

\begin{abstract}
  We investigate the initiation and formation of Coronal Mass
  Ejections (CMEs) via detailed two-viewpoint analysis of low corona
  observations of a relatively fast CME acquired by the SECCHI
  instruments aboard the STEREO mission. The event which occurred on
  January 2, 2008, was chosen because of several unique
  characteristics. It shows upward motions for at least four hours before
  the flare peak. Its speed and acceleration profiles exhibit a number
  of inflections which seem to have a direct counterpart in the GOES
  light curves. We detect and measure, in 3D, loops that collapse
  toward the erupting channel while the CME is increasing in size and
  accelerates. We suggest that these collapsing loops are our first
  evidence of magnetic evacuation behind the forming CME flux rope. We
  report the detection of a hot structure which becomes the core of
  the white light CME. We observe and measure unidirectional flows
  along the erupting filament channel which may be associated with the
  eruption process. Finally, we compare these observations to the
  predictions from the standard flare-CME model and find a very
  satisfactory agreement. We conclude that the standard flare-CME
  concept is a reliable representation of the initial stages of CMEs
  and that multi-viewpoint, high cadence EUV observations can be
  extremely useful in understanding the formation of CMEs.
   
\end{abstract}
\end{opening}

\section{Introduction} \label{sec:Introduction}

CMEs have been observed for more than 40
years now. They are one of the most energetic phenomena in our solar
system and the main driver of disturbances in the terrestrial space
environment. Despite observations of tens of thousands of CMEs, the
physical processes behind their formation and propagation have not yet
been understood completely
\cite{2001AGUGM.125..143K,2006SSRv..123..251F,2011LRSP....8....1C}. 

To make progress, we need to select the model (or models) that best
describe the phenomenon. To accomplish this, it is necessary to test
the theoretical predictions of the various models against the
observations as was discussed by
\inlinecite{2001AGUGM.125..143K}. Here, we concentrate on the
'standard' flare-CME model, also known as the CSHKP model
\cite{1992LNP...399....1S}. This is not actually a fully-fledged model
derived from the solution of a set of Magnetohydrodynamic (MHD)
equations but it is rather a two-dimensional (2D) cartoon
representation of the erupting process. However it captures the key
ingredients of many MHD models (i.e., the three-part CME, the ejection of
a flux rope, post-CME flaring loops, etc) and demonstrates, in a
straightforward way, the possible connection between the erupting and
flaring processes. For our discussion, we use the detailed model
representation in \inlinecite{2004ApJ...602..422L} (their Figure~1)
but many more variations can be found in the literature.

Even as a cartoon, the CSHKP model makes several predictions that can
be tested against the observations. First, it predicts the eruption of
a core surrounded by a cavity (or bubble) that forms during the
initiation process. High temperatures are expected in both the cavity
and the core as result of magnetic reconnection
\cite{2011LRSP....8....1C}. Second, the reconnection behind the
erupting system creates a magnetic void which draws adjacent lines
toward the current sheet thereby creating an inflow of material from
the surrounding flux systems. Third, through the reconnection
processes in the post-CME current sheet, magnetic energy is
transformed into thermal energy that powers the flare and kinetic
energy that powers the CME. Therefore, we expect a close
correspondence between the SXR light curve and the CME acceleration
profile as has been found in the past (e.g.,
\opencite{2004ApJ...604..420Z}). A delay between the two processes is
also likely depending on the magnetic fields and reconnection rates
involved \cite{2006ApJ...644..592R}. Fourth, there are many candidates
for the role of the eruption trigger. Flux emergence, tether-cutting
or even mass unloading from the prominence channel, are all capable of
driving the system out of its equilibrium state to set off the
eruption (see discussion and references in
\opencite{2011LRSP....8....1C}). Can the trigger be identified in
the observations?  

Many, if not all, of these predictions relate to the very first stages
of the CME; namely, its initiation and formation. However, the
initiation and formation stages of CMEs present some serious
observational challenges. The CME formation and initial evolution take
place low in the corona which is accessible only to imagers in the
Extreme Ultraviolet (EUV) or (less often) Soft X-Ray (SXR)
wavelengths. These instruments observe in a relatively narrow passband
and hence are sensitive to only a narrow range of temperatures, at a
time.  CME triggers, such as plasma instabilities occur within
Alfvenic temporal and spatial scales (of the order of tens of seconds
or hundreds of km for an active region). The subsequent energy release
also occurs in similar scales and the eruption is usually accompanied
by other phenomena such as flares, jets and lateral plasma motions
that may have nothing to do with the erupting structure but they
complicate the interpretation of the EUV observations.  

Therefore, we need observations of the
formation stages of a CME taken with high cadence and spatial
resolution but with minimal line-of-sight confusion. The unique
stereoscopic viewing and instrument complement provided by the
Sun-Earth Connection Coronal and Heliospheric Investigation
(SECCHI; \opencite{2008SSRv..136...67H}) on-board the {\it Solar
  TErrestial RElations Observatory (STEREO)}
 \cite{2008SSRv..136....5K} fulfills these requirements nicely.

To demonstrate this, we undertake a CME initiation study for an event
which took place on January 2, 2008. The eruption in the low corona
was observed very well by both SECCHI \textit{Extreme Ultraviolet Imagers
\/}(EUVI). We are able to examine in detail the various stages of the
initiation of a CME and relate them to the usual phenomena that
accompany these eruptions, such as flares and filament ejections. In
addition, we capture the transition of loop arcades into the forming
flux rope and report the first three-dimensional observations of loop
'implosion'. Taken together, these observations reveal many of the key
components of CME initiation and provide strong constraints for CME
models.  The paper is organized as follows. In Section~2, we present the
time history of the CME and discuss in detail several key observations
including the close correspondence between the acceleration profile
and the GOES SXR light curve, the novel observation and 3D
measurements of collapsing loops, the detection of a hot CME core, and
the observation of outflows along the filament channel. We
conclude in Section~3.

\section{Stereoscopic Observations of the January 2, 2008 CME}
\begin{figure}
\centerline{
\includegraphics[width=0.9\textwidth]{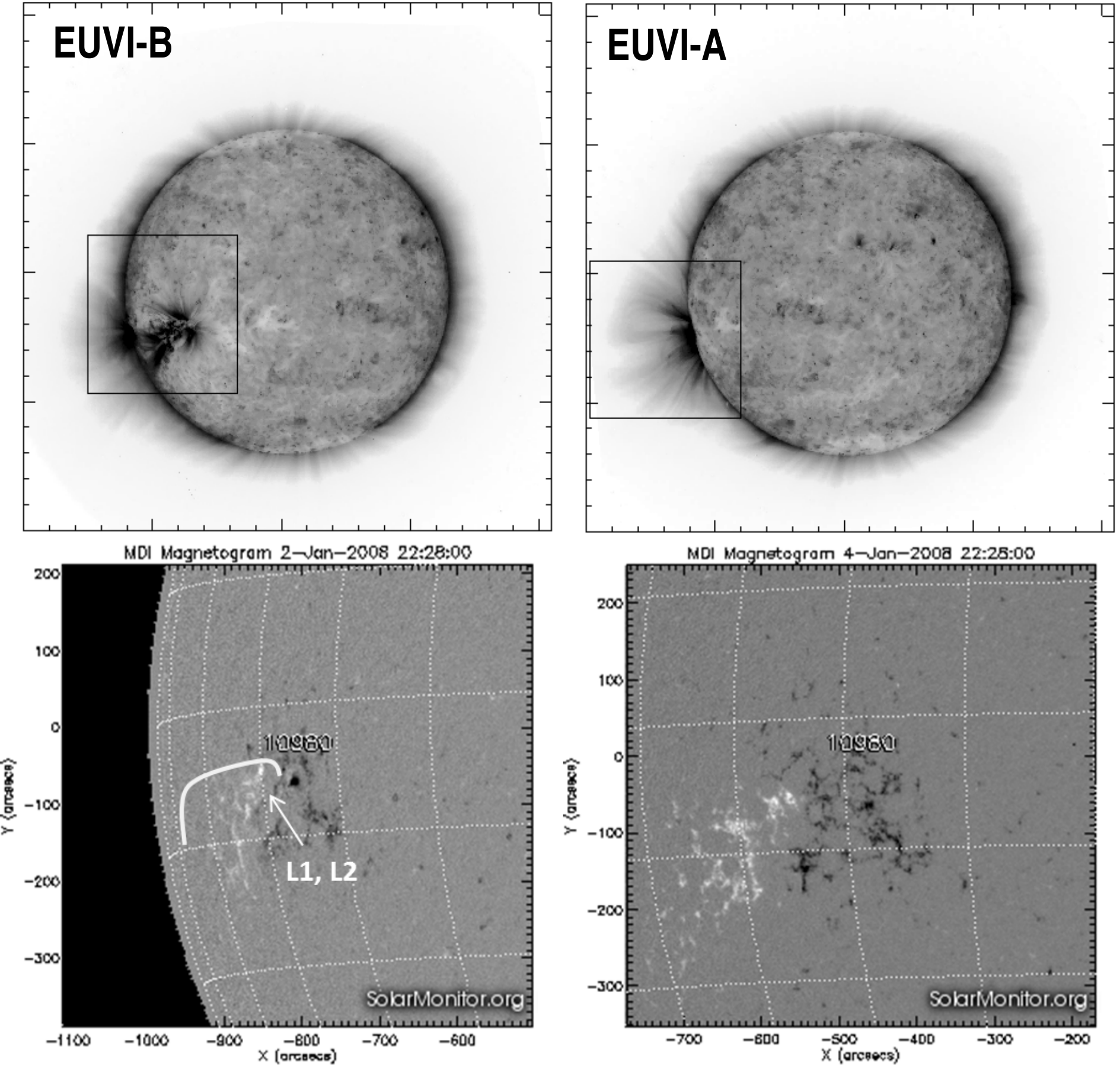}
}
\caption{Top panels: EUVI-A and -B 171\AA\ full disk images on January
  2, 2008 at 09:01~UT. The box marks the FOV around AR 10980 used in
  the subsequent analysis. Bottom panels: MDI magnetograms of AR10980
  on January 2 and 4 showing the magnetic field configuration for the
  event. The thick white line marks the filament channel involved in
  the eruption. The arrow mark the approximate location of collapsing
  loops discussed in \ref{sec:collapsing}. The magnetogram images are
  courtesy of SolarMonitor.org. }\label{fig:context}
\end{figure}

The event under study erupted from active region NOAA 10980 located at
S05E65 (Figure~\ref{fig:context}). The region has an alpha magnetic
configuration with a single leading negative polarity sunspot.  The
sunpot disappeared within a couple of days leaving only an extended
area of plage fields. The eruption occurred along a filament channel
(thick white line in Figure~\ref{fig:context}) overlying a neutral
line extending from the center of the region to its periphery.  The CME was
accompanied by a GOES C1.2 flare starting at 06:51 UT, peaking at
10:00~UT, and ending at 11:21~UT. It is therefore a long duration Soft
X-ray (SXR) event but the gradual rise of the light curve is also
indicative of a partially occulted event. Indeed, upward motions at
the location of the subsequent CME can be detected much earlier than
the flare peak as we shall see later. The event was observed by the
SECCHI/EUVI imagers on the STEREO-A and STEREO-B spacecraft which were
located $21^\circ$ West and $23^\circ$ East from the Sun-Earth line,
respectively. Therefore, it was a limb event for EUVI-A
($\sim88^\circ$) and an eastern event for EUVI-B ($\sim42^\circ$). The
3D kinematics of the CME in the SECCHI coronagraph fields of view have
been discussed in detail by \cite{Zhao10}.  Here, we focus on the
initiation of the CME in the low corona as witnessed in the EUVI
fields of view up to about 1.7 R$_{sun}$. We use mainly the 171\AA\
images because of their high cadence (150 sec) but we discuss the
observations in the other wavelengths as well. The images have been
processed by the \inlinecite{2008ApJ...674.1201S} wavelet-based
algorithm to enhance the visibility of the off-limb structures by
removing the instrumental stray light.

\subsection{The time history of the CME formation in the low corona}
Because of the unusual duration of the eruption, we have to find a
reliable marker for the start of the event. We use the time of the
first unambiguous detection of upward motion of EUV loops at the
location of the subsequent CME. This occurs at 06:13:30~UT (online
movie and Figure~\ref{fig:detail}). We rely on the EUVI-A images to
describe the upward evolution since the CME is propagating along the sky
plane of STEREO-A and therefore the images are least affected by
projection effects. 

The motion in EUVI-A originates in a high-lying loop system which
appears to encompass a cavity as evidenced by the lack of 171\AA\
emission (Figure~\ref{fig:detail}). Inside this cavity (in projection)
we detect a single bright loop (L1) that begins to collapse as the
rest of the loop system expands slowly. The loop is visible from
05:33:30~UT to 07:21:00~UT. The behavior of this collapsing loop is
almost immediately imitated by a larger loop arcade (L2). Their
collapse starts at around 08:18:30~UT. The CME front leaves the edge
of the EUVI-A field of view at 09:18:30~UT. The first evidence of a
CME core, in the traditional sense of a 3-part CME, becomes apparent
at 09:15:30~UT while the L2 system continues to collapse. At
10:03:30~UT, the loop arcade disappears, the CME continues to
accelerate and the usual post-eruptive arcade forms. An EUV wave is
launched by the expanding CME at around 10:33:30~UT. Material
continues to flow outward from the active region while the
post-eruptive arcade continues to grow until about 13:03:30~UT. We
take this time as the end of the eruption since it marks the end of
the material outflow and the growth of the flaring arcade.

The low-lying activity in the source region is not visible from EUVI-A
but it is clearly visible in EUVI-B. The images show that all the
action takes place along the filament channel running roughly
east-west through the center of the active region. The start of the
event occurs at the easternmost edge of the filament channel, closest to the
leading sunspot of 10890. The post-eruption loop system expands from
that location toward the east. The collapsing loops follow the same
path as they collapse (see online movie). The time history is summarized in
Table~\ref{tbl:history}.

\begin{table}
  \caption{Time history of the CME eruption as marked by several key
    events.} \label{tbl:history} 
\begin{tabular}{lcr}
Event                        & Time      & Elapsed time\\
                             &  (UT)     &  (min)\\
\hline
Upward motion (event starts) & 06:13:30	 & 0     \\
Single loop (L1) collapses   & 06:36:00	 & 22.5  \\
SXR flare starts             & 06:51:00  & 37.5  \\
Single loop (L1) disappears  & 07:21:00  & 67.5  \\
Loop arcade (L2) collapses   & 08:18:30	 & 125.0 \\
Core appears                 & 09:15:00  & 181.5 \\
Flaring Arcade (FL1) appears & 09:21:00	 & 207.5 \\
SXR Flare peaks              & 10:00:00  & 246.5 \\
Loop Arcade (L2) disappears  & 10:03:30	 & 230.0 \\
CME acceleration peaks	     & 10:23:00	 & 249.5 \\
EUV Wave appears             & 10:33:30  & 260.0 \\
End of SXR flare             & 11:21:00  & 307.5 \\
End of outflows (event ends) & 13:03:30  & 410.0 \\
\hline
\end{tabular}
\end{table}

\subsection{Height-Time Evolution of CME in the Low Corona} \label{sec:height}

Since the beginning of the day, the overlying loop system seems to be
in a steady state without noticeable motions other than the effect of
the solar rotation (the AR is rotating over the eastern limb as seen
from EUVI-A). Starting at around 6:13~UT, we can see upward motions
within the loop system and the whole system begins to expand after
6:31:30~UT. We choose to follow the top of the loops for our
height-time (ht) measurements. For the first two hours, however, the
motion is very slow and can be best appreciated by examining the
accompanying movie. Because of the slow rise, we use a running cadence
of 10 min (every four 171\AA\ frames) for the ht measurements to make the
motion easier to see.  Consequently, the height-time measurements were
taken with the full available cadence of 2.5 min. 

\begin{figure}
\centerline{\includegraphics[width=0.9\textwidth]{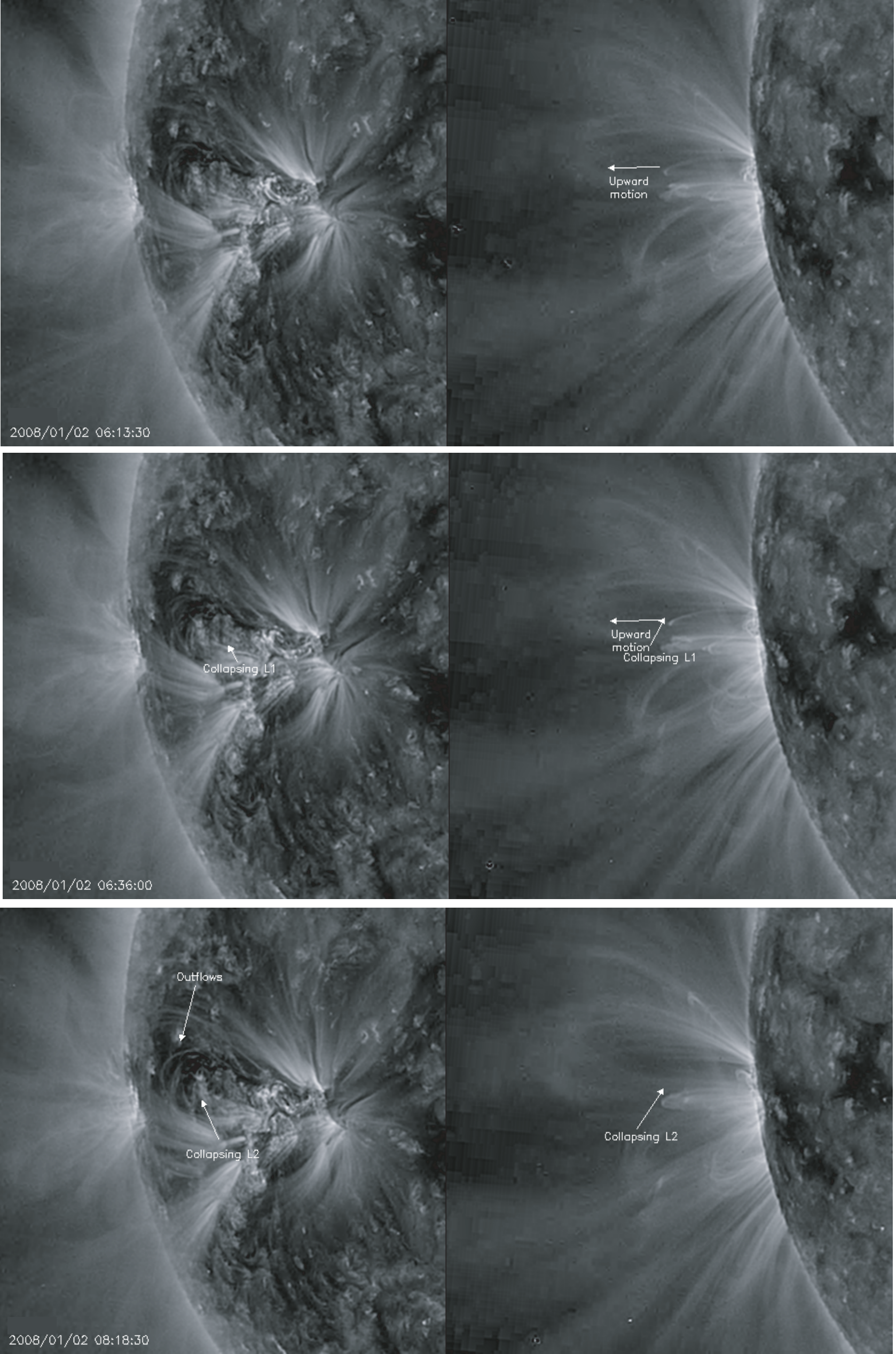}}

\caption{ Snapshots of the eruption as seen in simultaneous images
  from SECCHI/EUVI-A (right) and EUVI-B (left). The frames are taken
  from the online movie and the labeled features are discussed in
  Sections~\ref{sec:height} - \ref{sec:collapsing}. The times
  correspond to the EUVI-A observation time.  }\label{fig:detail}
\end{figure}

\begin{figure}
\centerline{\includegraphics[width=0.9\textwidth]{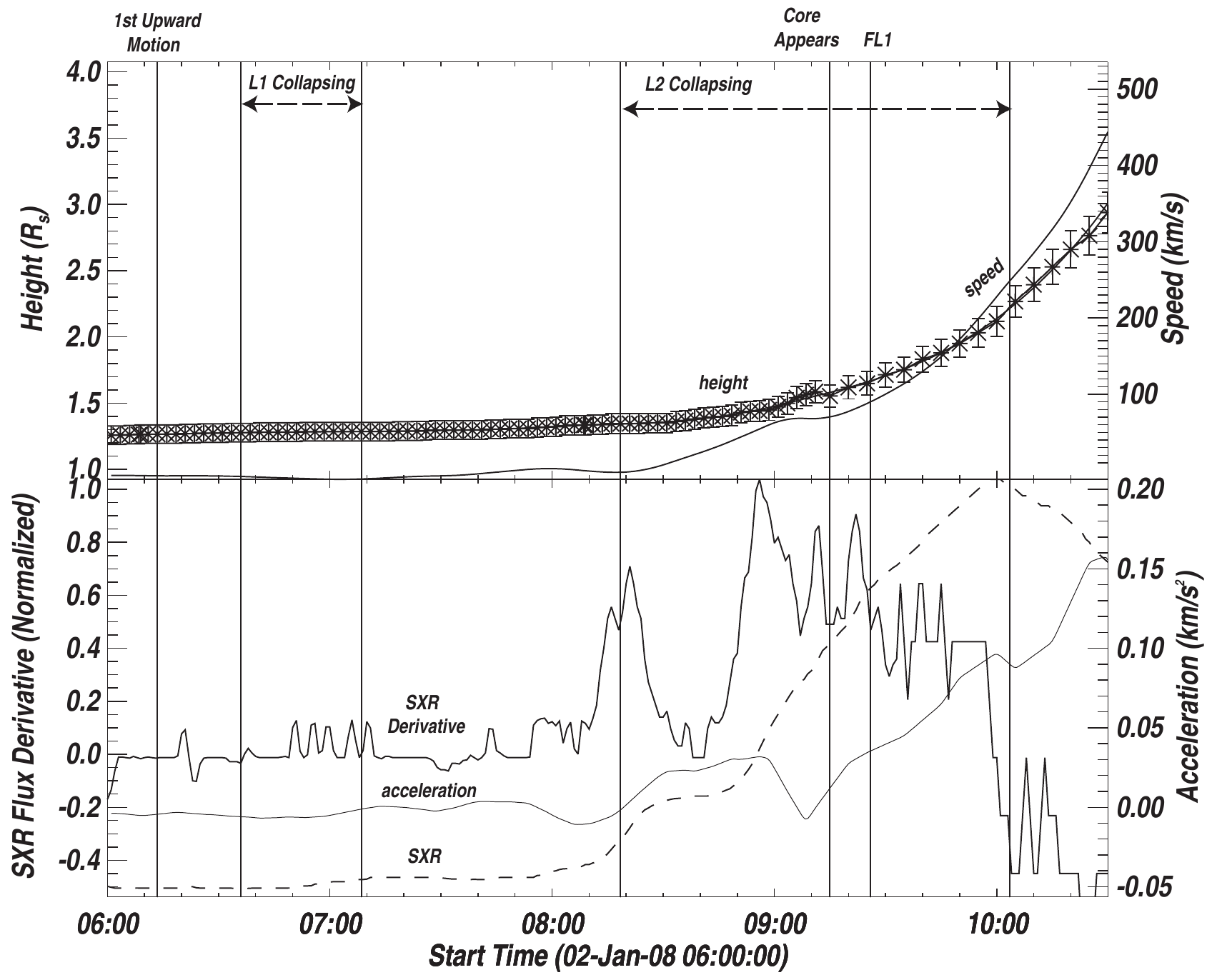}}
\caption{
  The development of the eruption as seen through height-time and velocity-time diagrams (top panel) and the Soft X-ray light curve and its derivative (bottom panel). The heights correspond to the top of the CME structure and the speed is derived using a smoothing procedure (Section~2.2). Key events, such as collapsing loops, are also marked on the figure and are discussed in Sections~2.3 and 2.4.
        }\label{fig:CME_height} 
\end{figure}

EUVI-A is able to follow the loop top until 09:18~UT, when the CME
exits the telescope's field of view (Figure \ref{fig:detail}). We then
turn to the COR1-A images to obtain a complete set of ht measurements
during the rise time of the SXR flare. The measurements are presented
in Figure \ref{fig:CME_height}. The first COR1-A ht point is plotted
right next to the line labeled 'Core Appears'. We did not attempt to
triangulate the CME front positions in EUVI-B because the front is
visible only between 8:51 - 9:08~UT and is quite extended and
diffuse. The projection, however, does not affect our EUVI-A
measurements because it is clear that the CME lies very close to the
EUVI-A plane of the sky. To derive velocity and acceleration profiles
from our sparse ht points, it is always better to smooth the ht points
first. We use the same smoothing method as in
\inlinecite{2010ApJ...724L.188P}. Namely, we minimize the $\chi^2$ between
the data and a cubic spline plus a penalty function equal to the
second derivative of the spline multiplied by a weighting factor,
$spar$, provided by the user. In this case, $spar = 0.6$ offers the best balance
between noisy and overly smooth acceleration profiles. The results are
shown in Figure~\ref{fig:CME_height} where the velocity is plotted in the top
panel and the acceleration in the bottom panel.

The last height measurement was taken in COR1-A at 10:50:18~UT but we
show results only until 10:30~UT. At that point the CME has reached a
height of $3 R_{\odot}$ with velocity of $420$ km s$^{-1}$. Both our
speed and acceleration results are consistent with the
\inlinecite{Zhao10} results which were based on ht measurements after
10:00:00 UT and on a different technique. 

In the bottom panel of Figure~\ref{fig:CME_height}, we compare the CME
acceleration to the 1-min GOES SXR light curve (1-8\AA\ channel) and
its time derivative which is considered a proxy to energy release
episodes. Both SXR curves are normalized to their respective
peaks. 

First, we see that the CME acceleration profile follows closely the
SXR rise as seen before
\cite{2004ApJ...604..420Z,2008ApJ...673L..95T,2010ApJ...712.1410T},
albeit with some time delay. This delay is consistent with the gradual
character of this CME. Generally speaking, impulsive CMEs tend to have
acceleration profiles leading the SXR flux profile
\cite{2010ApJ...724L.188P} since it takes some time to heat the
chromosphere and to fill in the coronal loops with the hot plasma. In
our case, the CME acceleration peaks sharply after at about 10:23~UT when
the flux rope core and the post-CME flaring arcades appear. We return
to this point in Section~\ref{sec:flux-rope}.

Second, the impulsive phase of flare is a bit unusual because the rise
of the SXR flux is marked by two interim inflections (one at
$\sim$8:25-8:50 and the second at 9:20~UT) before the SXR peak at
10:00~UT. Remarkably, the CME acceleration profile changes at almost
the same times. We can discern inflection points at approximately
8:30, 8:55, 9:10, 9:20, 9:55, and 10:25~UT in the bottom panel of
Figure~\ref{fig:CME_height}. These points bracket intensity changes in
the SXR light curve and coincide with peaks in the SXR derivative (and
hence energy release episodes). The correlations are positive
(acceleration) with the exception of the SXR derivative peak at
9:10~UT which occurs during a decelerating phase of the CME. The time
offsets between the SXR and CME acceleration peaks are within 5 min of
each other. There is even indication for an earlier acceleration jump
associated with a small step in the SXR flux at around 7:00~UT. Since
flaring and hence changes in the SXR profile result from energy
release in the low corona, it is tempting to interpret the changes in
the CME acceleration profile as a result of the same energy
release. For example, the CME speed increases from about 5 km s$^{-1}$
to almost 80 km s$^{-1}$ during the first flaring episode, between
8:20 and 9:00~UT. To investigate whether the correspondence between
the SXR and CME acceleration profiles is based on a causal
relationship we look into the various phases of the event in detail in
the following.

\subsection{Collapsing Loops} \label{sec:collapsing}

The observation of the two collapsing loop systems, L1 and L2,
represents a unique aspect of this event and drew our attention to
it. The first system, L1, appears to be a single loop which stands out
because it is projected against an area of reduced 171\AA\ emission,
possibly a cavity, as viewed from EUVI-A. The loop appears to collapse
starting at around 6:36~UT and disappears at 7:21~UT. The loops do
not appear to simply contract as has been seen in other occasions (see
\opencite{2011SSRv..158....5H} and references therein) but it rather
seems to incline toward the cavity. At the same time, the cavity is
slowly rising and expanding. This behavior, especially the
disappearance of the loop, is seen for the first time and is
suggestive of a magnetic relationship between the loop and the
cavity. But before we discuss this further, we have to understand the
3D topology of the loop.

\begin{figure}
\centerline{\includegraphics[width=0.7\textwidth]{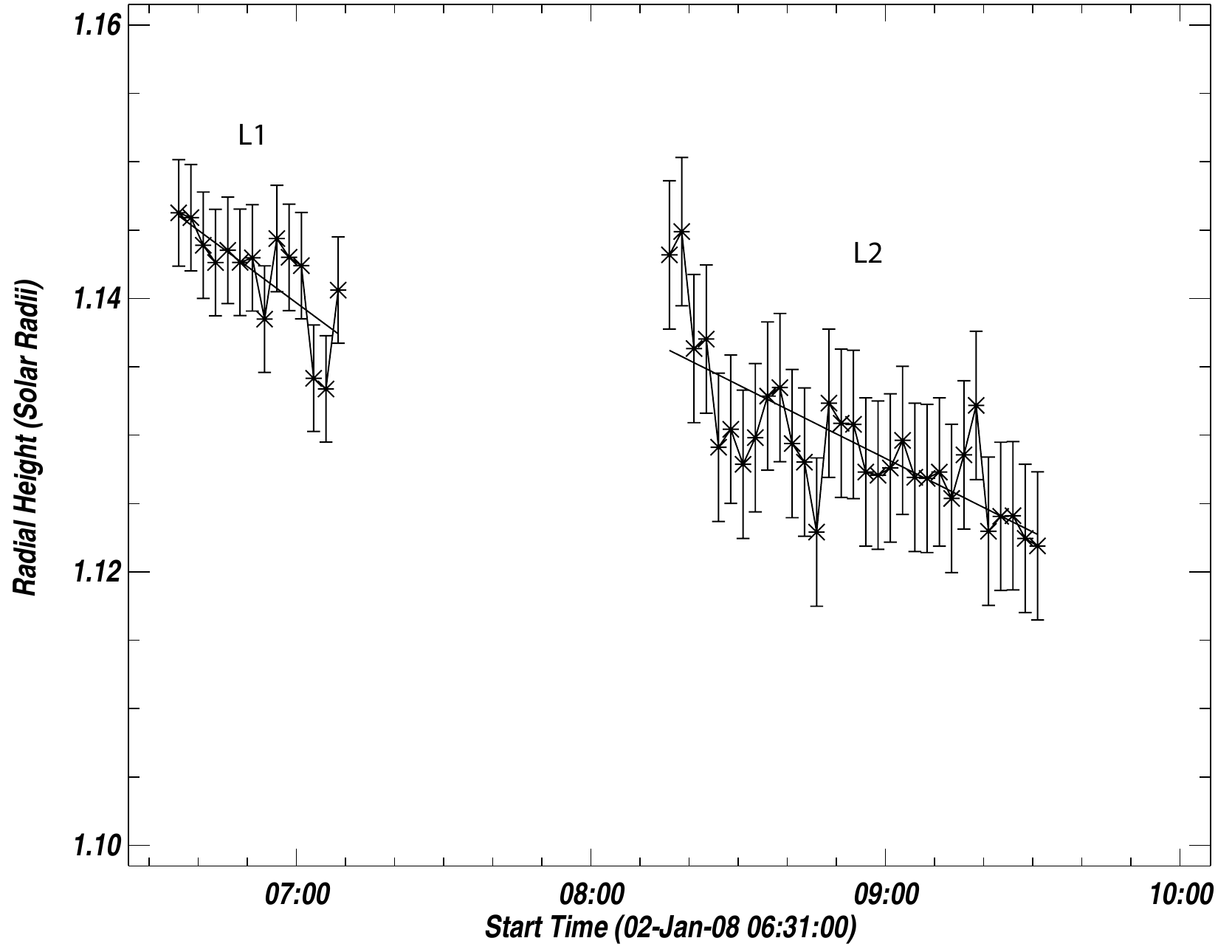}}
\caption{Height-time measurements of the two sets of collapsing loops
  observed during this CME event. The heights are true radial
  distances obtained via triangulation of the structures in the EUVI-A
and -B 171\AA\ images. The solid lines represent linear fits to the ht
points and result in speeds of 3 km s$^{-1}$ and 2 km s$^{-1}$ for the
L1 and L2 systems, respectively.} \label{fig:collapsing-loops-ht}
\end{figure}

The loop is quite tall (0.15 R$_\odot$ or $1.04\times 10^5$
km). However, it is very hard to discern from the EUVI-B perspective
because it is narrow (small footpoint distance) and is oriented
toward EUVI-B (Figure~\ref{fig:detail}, middle panels). Nevertheless,
its 3-dimensional (3D) orientation can be established because it
becomes visible in EUVI-B once it starts collapsing. We use standard
SECCHI software (the \textit{scc\_measure\/} routine) to derive its 3D
parameters as a function of time for the period 6:36-7:08~UT. Briefly,
the algorithm requires the user to select a point in the loop in one
view. This selection corresponds to a line (the epipolar line) in the
other view.  The successful triangulation is achieved by identifying
the location where the epipolar line intersects the projection of the
original point in the loop. In our case, the obvious candidate is the
bright loop-top in EUVI-A. Unfortunately it does not have a clearly
identifiable counterpart in EUVI-B because we view the loop-top
face-on. After careful examination of the movies, we decided to use a
relatively bright edge in EUVI-B as the starting point because it was
easier to find the intersection of the epipolar line with the loop in
the EUVI-A images. The intersection was located a few pixels below the
bright loop apex along the loop leg farthest from the EUVI-A
observer. Here, we are primarily interested in the temporal behavior
of the loop height. The ht measurements are shown in the
Figure~\ref{fig:collapsing-loops-ht}. There is an obvious downward
trend despite some scatter in the measurements around 7~UT. The
scatter arises from inaccurate identification of the same part of the
structure in the two images. We repeated the measurements three times
but we were not able to improve the scatter in time. Although the
scatter in the three measurements (at the same time) was very small,
we decided to adopt a conservative error estimate equal to the
standard deviation of all measurements in order to account for the
scatter in time. Given the scatter, we fit the ht data points with a
first order polynomial, assuming therefore, a constant speed. We
obtained a speed of $3$ km s$^{-1}$.

\begin{figure}
\centerline{\includegraphics[width=\textwidth]{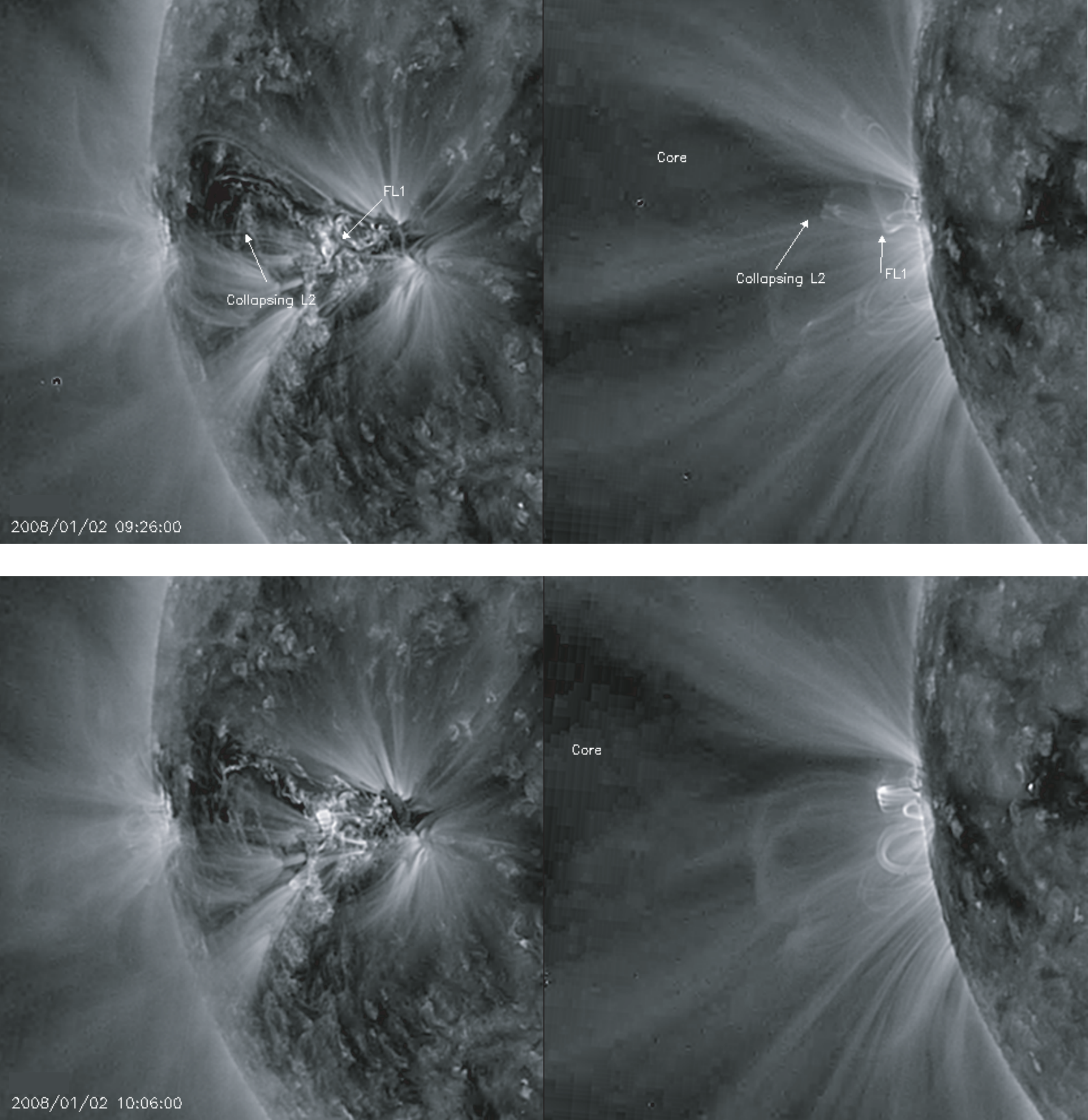}}
\caption{The collapsing loops toward the expanding CME cavity as seen from EUVI-A
  (top right) and EUVI-B (top left). The arrows point to the direction of the collapse. A flaring loop system with peculiar connectivity is also marked (FL1). The bottom panels show snapshots at the time of the disappearance of the L2 system and the appearance of bright flaring loops at their footpoints.} \label{fig:infalling-loops}
\end{figure}

Just an hour later, at 8:18~UT, a larger loop system (L2) begins to
collapse following an almost identical path to L1
(Figure~\ref{fig:infalling-loops}). The L2 system is located just a few
pixels southeast of L1 and reaches almost the same height, 0.15
R$_\odot$. L2 is more discernible in the EUVI-B images but it could
easily be overlooked if it was not for the EUVI-A
observations. This is a very important point and explains the lack of
such observations in the past. How many times have we missed such
inclining, collapsing loops in the past because we had only one
viewpoint available? Thanks to the two EUVI views, we can derive the
3D orientation of L2 as we did for L1. The resulting ht points in
Figure~\ref{fig:collapsing-loops-ht} show a rather sharp drop in the first
15 mins followed by a gradual contraction. We chose to fit again a
first order polynomial to describe the long-term evolution of the loop
apex. In this case, we derived a slight slower speed of $2$ km
s$^{-1}$. The L2 system collapses toward the bottom of the erupting structure and the cavity
is clearly rising while the loop system is collapsing. The loops
disappear similarly to L1, at a height of 0.12 R$_\odot$. We note that
the CME clearly took off while the L2 system was still collapsing and
that the disappearance of the L2 loops coincides with the flare
peak. It is also worth noting (Figure~\ref{fig:detail}) that the first
set of bright flaring loops (in 171\AA) appears at the location of the L2
footpoints.

The coincidence of the collapsing loops to the rise and growth of the
erupting structure is very suggestive of a magnetic connection between
the two and is expected according to the standard CME
models. Specifically, the models show that as the flux rope rises and
a current sheet forms behind it, the resulting reconnection attracts
nearby magnetic lines. The result is the creation of a void which
field lines further afield would rush to fill. The void, and
subsequent inflow, would occur across the erupting channel. Because
most models are essentially two-dimensional, the reconnection is
symmetric and proceeds from the center of the neutral line (or
filament channel) outwards and across the channel. In this situation,
the inflows are depicted on either side of the post-CME current sheet
(e.g., \opencite{2004ApJ...602..422L}). However, this does not have to
be, and most likely it is not the situation with the actual
observations. Erupting prominences (a usual proxy for the CME core)
are often seen rising asymmetrically and the majority of H$\alpha$
ribbons brighten progressively both across and along the channel. If
the eruption were to start at one end of the filament channel then the
ribbons would move from that end of the channel to the opposite
instead from starting at the middle and propagate outwards along the
channel as the symmetric picture would suggest
\cite{2009ApJ...690..347L}. In that case, the void would form on end
of the channel and any likely inflows would occur there. Such an
asymmetric eruption was discussed by
\inlinecite{Patsourakos_Vourlidas_Kliem_2010}. Therefore, we expect
the following: (i) inflows toward and behind the erupting structure, (ii) the
inflows would occur where the flux rope rises first, and (iii) the
inflows and flux rope growth would be correlated. The analysis of the
collapsing loops meets all three of these expectations and hence we
claim that they constitute the first direct evidence of the process of
flux rope formation (or growth) though the incorporation of
neighboring flux systems into the erupting structure.

\subsection{The Detection of the Hot Flux Rope Core} \label{sec:flux-rope}
\begin{figure}
\centerline{\includegraphics[width=\textwidth]{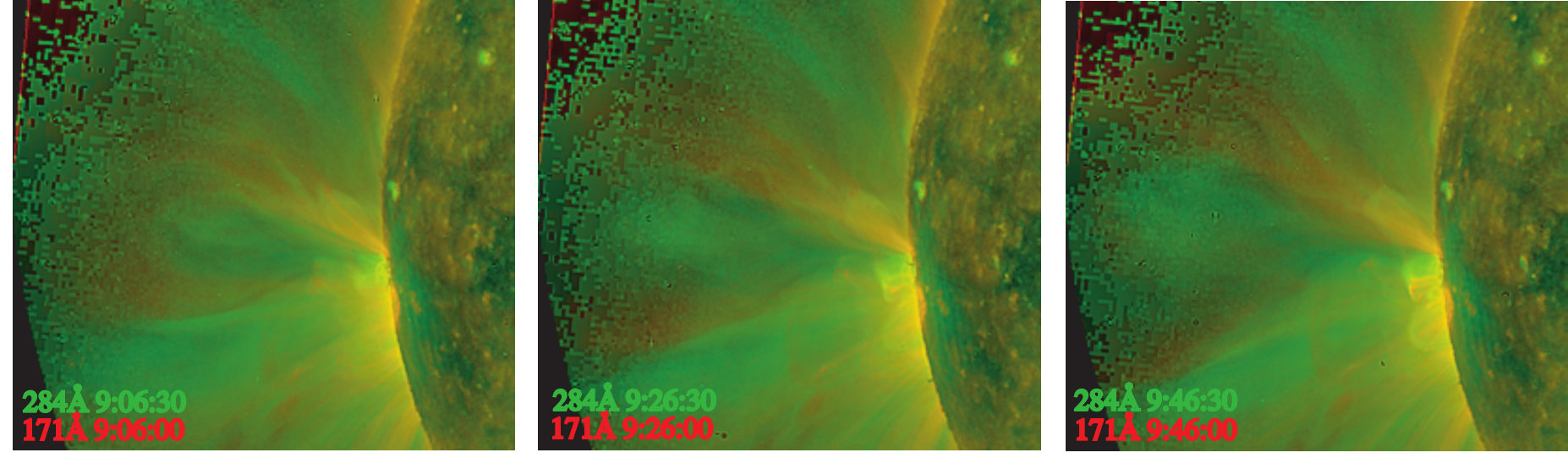}}
\caption{Overlays of quasi-simultaneous EUVI-A observations at 284\AA\
  (green) and 171\AA\ (red) during the appearance of the CME core. The
degree of color dominance (green or red) at a given location can be
used as a proxy for the temperature of the material at that
location. For example, the CME core appears fully green at 9:26~UT
which implies that most of the core material is emitting at 284\AA\ or
about 1.8 MK, at that time.}\label{fig:fluxrope}
\end{figure}
The CME has a clear 3-part structure in the COR1 and COR2 observations
\cite{Zhao10} and both the front and following cavity are easily
discernible in the 171\AA\ observations. The counterpart for the core
is not easy to identify until 9:15~UT when a rather diffuse blob-like
structure appears in the 195\AA\ images. No erupting prominence is
detected in the 304\AA\ observations. The core is clearly visible in
the 284\AA\ image taken at 9:26:30~UT but it is very hard to detect in
the almost simultaneous 171\AA\ image at 9:26~UT
(Figure~\ref{fig:fluxrope}). The dominant contribution in the 284\AA\
bandpass comes from the FeXV line which forms at around 1.8
MK. Therefore, the lack of 171\AA\ emission and the bright 284\AA\
emission suggest that the majority of the core plasma comes from hot
temperatures. This is exactly what the models predict and recent \textit{Solar
Dynamics Observatory\/} (SDO)
observations show \cite{2011ApJ...732L..25C}.  Therefore, we conclude
that the CME core in our event is hot and comes at the tail end of the
cavity within the erupting structure. Once the core is identified in
the 284\AA\ and 194\AA\ images, it is relatively straightforward to
follow in the 171\AA\ as well although it remains quite faint (see
online movie).

\subsection{Flows along the Filament Channel}

\begin{figure}
\centerline{
	    \includegraphics[height=0.5\textwidth, angle=90]{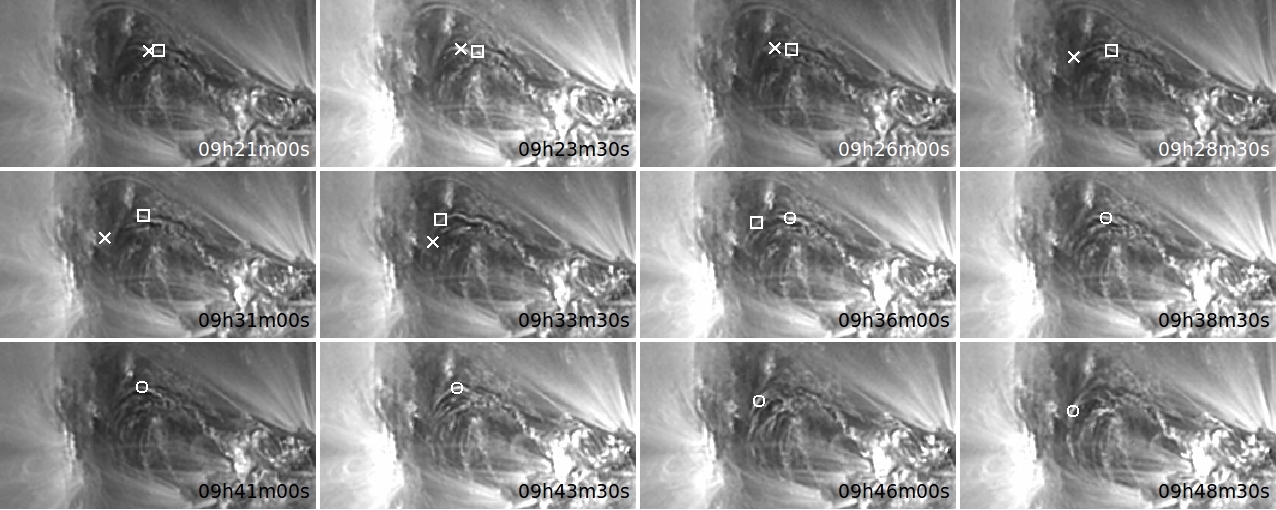}
	    }
            \caption{Flows along the erupting filament channel. The
              symbols indicate a particular blob tracked at different
              times in each of these EUVI-B 171\AA\ images.
            }  \label{fig:siphon}
\end{figure}

\begin{figure}
\centerline{
	    \includegraphics[width=\textwidth]{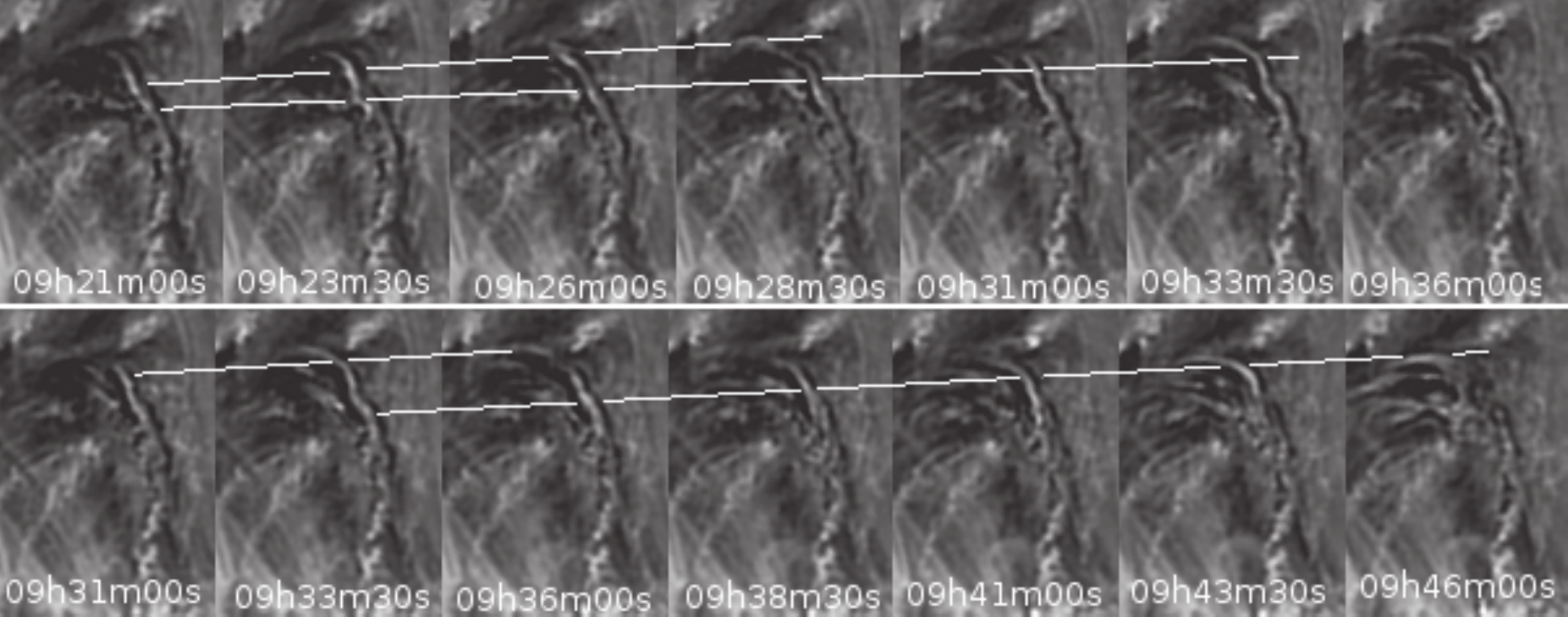}
	    }
            \caption{Demonstration of our tracking of the blobs in the
              EUVI-B images. The area was rotated to make easier the
              display of lines connecting the blobs. In the top panels,
              the upper line traces the blob marked with an X
              in Figure~\ref{fig:siphon}, and the lower line traces the
              blob marked with a box. In the bottom panels, the first
              two frames are repeats from the last two frames of the
              top panel. The upper line is the continuation of the
              trace for the blob marked with the box symbol, and the
              lower line traces the blob marked with the circle.
            } \label{fig:blobs}
\end{figure}

Throughout the event, one can observe flows along the filament channel
(FC). They become more obvious along a bend of the FC at its eastern
end. The filament itself is observed as a collection of dark threads
in the 171\AA\ channel due to the absorption from the cool
material. It is anchored in the AR on its western end and in the quiet
sun at its eastern end. The flows seem to evolve in two phases. In the
first one, which lasts until 8:28~UT, the flows are brighter. In the
second phase, which lasts until 10:06~UT, the flowing material
acquires a blob-like character. Some of those blobs are depicted in
Figure \ref{fig:siphon}. The symbols in this figure (cross, box,
circle) indicate the position the blobs we identified and measured at
different time frames. In Figure \ref{fig:blobs}, the area of interest
has been rotated to make the blob movement more obvious. The position
of each blob in this sequence of images is connected with a line.

After tracing the blobs, we measured their velocities. When the size
of blobs was small (e.g. at 09:23:00 UT), their position was assumed
to be their coordinates in the image. When the blobs became more
extended (e.g. at 09:38:30 UT), we took the middle point as their
average position, and their length was taken as the error uncertainty.

Because the blobs were located very low in the corona, they were not
visible from EUVI-A. Because we know the angular distance of EUVI-A
and the location of the flows from EUVI-B, we can derive an upper
limit for the height of the channel of 0.015 R$_\odot$ or
$10.5\times10^4$ km. Since they move parallel to the surface and over
a limited spatial extension, there was no need to correct for
spherical geometry. The effect is less than 4\% for the full
$30^\circ$ length of the filament which we did not use in our
measurements. However, the projection effect due to the proximity of
the channel to the limb needs to be taken into account. The flows are
measured at about $65^\circ$ east longitude so the correction factor
is $\sim 1/cos(65^\circ) \sim 2.36$. The average deprojected
velocities of the blobs are given in Table \ref{tab:siphon}. Each blob
is named after the symbol we used to mark them in
Figure~\ref{fig:siphon}.

The relation of the flows to the eruption is not immediately
clear. First, they appear to correspond to material flowing out of the
AR into the quiet sun because they propagate only in one direction,
from the center of the AR toward the quiet sun. Such behavior has been
very common since the beginning of EUVI observations and is always
related to AR filaments that extend into the quiet sun. Examples can
be seen in the eruptions of 1, 16, and 19 May 2009, 5 and 9 April
2008, 14 and 18 August 2010. The event on 3 April 2010 has been
analyzed in the detail by \inlinecite{2011ApJ...727L..10S} who connect
 such flows to off-loading of cool plasma that may contributed to
the subsequent CME eruption. Second, the nature of the blobs changes
at around 8:28~UT from thick elongated flows to smaller blob-like
features suggesting that the amount of the flowing material has been
reduced or the plasma has cooled down. It is interesting to note that the CME underwent its first
acceleration jump during that time. This apparent correlation seems to
support the \inlinecite{2011ApJ...727L..10S} interpretation of the flows as
off-loading material and suggests that gravity may affect the early
acceleration profile of CMEs.
\begin{table}
  \caption{Average velocities for each of the three blobs.  The names
    correspond to the symbols used to mark the blobs in Figure
    \ref{fig:siphon}.} 
  \begin{tabular}{lrr}
    \hline
    Blob 	&  Velocity $(km\, s^{-1})$	&	Error \\
    \hline
    X		&	125		&	5.3 \\
    Box		&	116		&	4.9 \\
    Circle	&	130		&	5.4 \\
    \hline
  \end{tabular}
\label{tab:siphon}
\end{table}
There is an alternative explanation, however. The flows apparently
trace closed field lines along the filament. The movement of the blobs
is directed away from the site of the emerging fluxrope where energy
input is taking place leading to higher plasma pressures in its vicinity. Therefore, the observed flows could be siphon
flow imposed by a pressure difference between the two footpoints of the
filament \cite{Cargill_Priest_1980,Cargill_Priest_1982}. 

\section{Discussion and Conclusions}

We investigate in detail the initiation and formation of a CME on
January 2, 2008 using two-viewpoint EUV observations in the lower
corona. The images are obtained in the 171\AA\ (150 sec cadence) and
284\AA\ (20 min cadence) channels of the EUVI instruments aboard the
{\it STEREO} mission. The event evolves slowly for several hours but
it then quickly accelerates around the time of the accompanying SXR
flare. This allows us to study in detail both its evolution toward the
eruption, the subsequent formation of a CME, and its connection to the
flare energy release profile. Our main results can be summarized as
follows:
\begin{itemize}
\item The acceleration profile of the CME is quite variable with peaks and
  valleys. The acceleration changes are similar, in time of appearance and duration, with
  corresponding changes in the GOES SXR light curve. 
\item The CME acceleration peaks at 10:30~UT which is 30 mins after
  the peak of the SXR flare.
\item The upward motions of the (eventually) erupting structure
  started at 6:13 UT, about 1 hour before a small SXR flux increase
  and 2 hours before a significant increase of SXR flux occurred
  (Figure~\ref{fig:CME_height}).
\item We detect, for the first time, two sets of collapsing loops. The
  two viewpoint EUVI observations allow us to measure their 3D
  evolution. They shrink very little (compared to past observations of
  shrinking loops) so most of their collapse is due to their inclining
  toward the erupting channel, beneath the rising cavity. They appear
  in all EUVI channels and they disappear in all of them at a height
  of 0.12 R$_\odot$. The post-CME arcades appear after the
  disappearance of the collapsing loops and at the same location. The
  CME cavity is clearly growing while the second loop system (L2) is
  collapsing. These observations lead us to conclude that the two loop
  systems are likely drawn behind the expanding magnetic cavity
  surrounding the CME core. This appears to be the first detection of
  this process predicted by CME initiation models.
\item We detect the core of the CME mostly in the hot EUVI channel at 284\AA\
  (1.8 MK) and the 195\AA\ channel. This observation provides further
  support that the CME cavity contains hot plasma as recent AIA
  observations have shown \cite{2011ApJ...732L..25C}.
\item We detect significant and long duration ($\sim3$ hours) plasma flows
  along the filament channel before its eruption. Their nature changes
  abruptly at around 8:30~UT coincident with a sudden change in the
  rising speed of the cavity. This coincidence suggests that
  mass unloading is perhaps playing a role in the early CME
  kinematics. 
\item The direction of the flows, from the western to the
  eastern part of the active region, is also in agreement with the
  temporal evolution of the flaring ribbons and post-eruptive flaring
  arcades, and the direction of the
  collapsing loops. Clearly, the eruption starts at the center of the
  active region and propagates to the east along the filament channel
  and toward the quiet sun footpoints of that channel. 
\item Despite the large number of novel observations and detailed
  measurements we cannot tell with certainty whether the erupted
  flux rope was pre-existing or was formed during the
  eruption. However, we are fairly certain that additional flux was
  introduced in the erupting flux rope during its ascent. This is the
  second event we reach this conclusion
  \cite{Patsourakos_Vourlidas_Kliem_2010} and is the expected outcome
  of several models
  \cite{2004ApJ...602..422L,2006SSRv..123..251F,2011LRSP....8....1C}. It
  is, therefore, important to take this effect into account in the
  estimation of magnetic flux entrained in CMEs.
\end{itemize}
All these observations confirm corresponding expectations of the
standard flare-CME models and suggest that such models are likely
reliable representations of the eruption process in the corona. Our
analysis demonstrates the power of two-viewpoint observations of the
low corona and the importance of extended fields of view for EUV
instruments so that the acceleration profile of the CME and the
relationships among the various erupting
structures can be measured consistently.  
\begin{acks}
  We thank the referee for the very useful comments and G. Stenborg
  for providing the wavelet-enhanced EUVI images and S. Patsourakos
  for fruitful discussions. The work of AV is supported by NASA
  contract S-136361-Y to the Naval Research Laboratory. The SECCHI
  data are produced by an international consortium of the NRL, LMSAL
  and NASA GSFC (USA), RAL and Univ. Bham (UK), MPS (Germany), CSL
  (Belgium), IOTA and IAS (France).
\end{acks}
\bibliographystyle{spr-mp-sola-cnd}

\bibliography{cme}


\end{article}

\end{document}